# Title: Elasto-inertial turbulence

Devranjan Samanta[1,2], Yves Dubief[3], Markus Holzner[1†], Christof Schäfer[2], Alexander Morozov[4], Christian Wagner[2] and Björn Hof [1*]

**Affiliations:**

[1]Max-Planck-Institute for Dynamics and Self-Organization, Bunsenstrasse 10, 37073 Göttingen, Germany.

[2] Experimental Physics, Saarland University, Saarbrücken, Germany.

[3] School of Engineering, University of Vermont, Burlington, VT 05405, USA.

[4]School of Physics & Astronomy , University of Edinburgh, Mayfield Road, Edinburgh EH9 3JZ, UK.

*Correspondence to: bhof@gwdg.de

†current address: Institute of Environmental Engineering, ETH Zürich, Switzerland.

**Abstract**:

**Turbulence is ubiquitous in nature yet even for the case of ordinary Newtonian fluids like water our understanding of this phenomenon is limited. Many liquids of practical importance however are more complicated (e.g. blood, polymer melts or paints), they exhibit elastic as well as viscous characteristics and the relation between stress and strain is nonlinear. We here demonstrate for a model system of such complex fluids that at high shear rates turbulence is not simply modified as previously believed but it is suppressed and replaced by a new type of disordered motion, elasto-inertial turbulence (EIT). EIT is found to occur at much lower Reynolds numbers than Newtonian turbulence and the dynamical properties differ significantly. In particular the drag is strongly reduced and the observed friction scaling resolves a longstanding puzzle in non-Newtonian fluid mechanics regarding the nature of the so-called maximum drag reduction asymptote. Theoretical considerations imply that EIT will arise in complex fluids if the extensional viscosity is sufficiently large.**

The most efficient method to reduce the large drag of turbulent flows of liquids is through addition of small amounts of polymers or surfactants. As first observed in the 1940's (1) frictional losses can be reduced by more than 70% (2,3) and this technique has found application in oil pipelines, sewage, heating and irrigation networks (4,5). For dilute solutions the drag is found to reduce with polymer concentration and eventually approaches an empirically found limit, the maximum drag reduction (MDR) asymptote (6). In the drag reduction process the elasticity of the long chain polymer molecules plays a decisive role. The molecules are stretched in strong shear and elongational flow and recoil in vortical regions. It has been shown that this process inhibits vortices and hence suppresses the turbulence sustaining mechanism (7,8). While a variety of theories have been put forward to explain the details of the underlying mechanisms (3) they all interpret the resulting flow (in particular in the asymptotic limit) as a modified form of ordinary shear flow turbulence.

Other studies of polymer solutions showed that in the limit of small Reynolds number (ratio of inertial to viscous forces) and large Weissenberg number (Wi=$\lambda\gamma$ is the product of the longest polymer relaxation time $\lambda$ and the shear rate $\gamma$) a new type of disordered motion, called elastic turbulence exists (9) following an elastic instability (10,11). Instead of inertia here elastic stresses destabilize the flow and sustain spatio-temporal disorder. A linear instability of this kind, however, only occurs in flows with curved streamlines, and it hence cannot be responsible for the phenomena observed in this paper. A different destabilizing effect of viscoelastic fluids has been reported in some early studies of polymeric solutions (12,13,14). Here it was found that for sufficiently large polymer concentrations, turbulence can set in at lower Reynolds numbers than in the Newtonian case and this effect has been termed early turbulence. In the following we will show for the case of pipe flow that this viscoelastic instability is not limited to high polymer concentrations but will generally occur if Re is sufficiently large. Furthermore we show that for high shear rates (large Wi) this state of motion subdues and replaces ordinary turbulence and hence dominates the dynamics.

**Results.** At the lowest Reynolds numbers where turbulence is sustainable in pipe flow it appears (18) in the form of axially localized structures about 20 D in length, so-called turbulent 'puffs'. It has been shown that these structures decay back to laminar after sufficiently long times following a memoryless process (16,19). Hence for each Reynolds number there is a distinct probability that a turbulent puff will survive beyond a certain time horizon. In the first set of measurements this characteristic was used to quantify the influence of polymers on the transition to turbulence. Experiments were carried out in pipe flows for various different polymer concentrations (50ppm, 100ppm, 125ppm,150ppm and 175ppm). In all cases the survival probability of puffs increases with Reynolds number and owing to the transient nature of the turbulent puffs is found to only approach a probability of one asymptotically with Re (Fig. 1A). When compared to pure water (blue curve) the curves are shifted to larger Re as the concentration is increased showing that the polymers delay transition and subdue turbulence. The Reynolds number required to reach a P=0.5 survival probability is found to increase faster than linear (Fig. 1B) with polymer concentration, providing a measure of the rate at which the turbulent state is postponed to larger Re (i.e. transition delay).

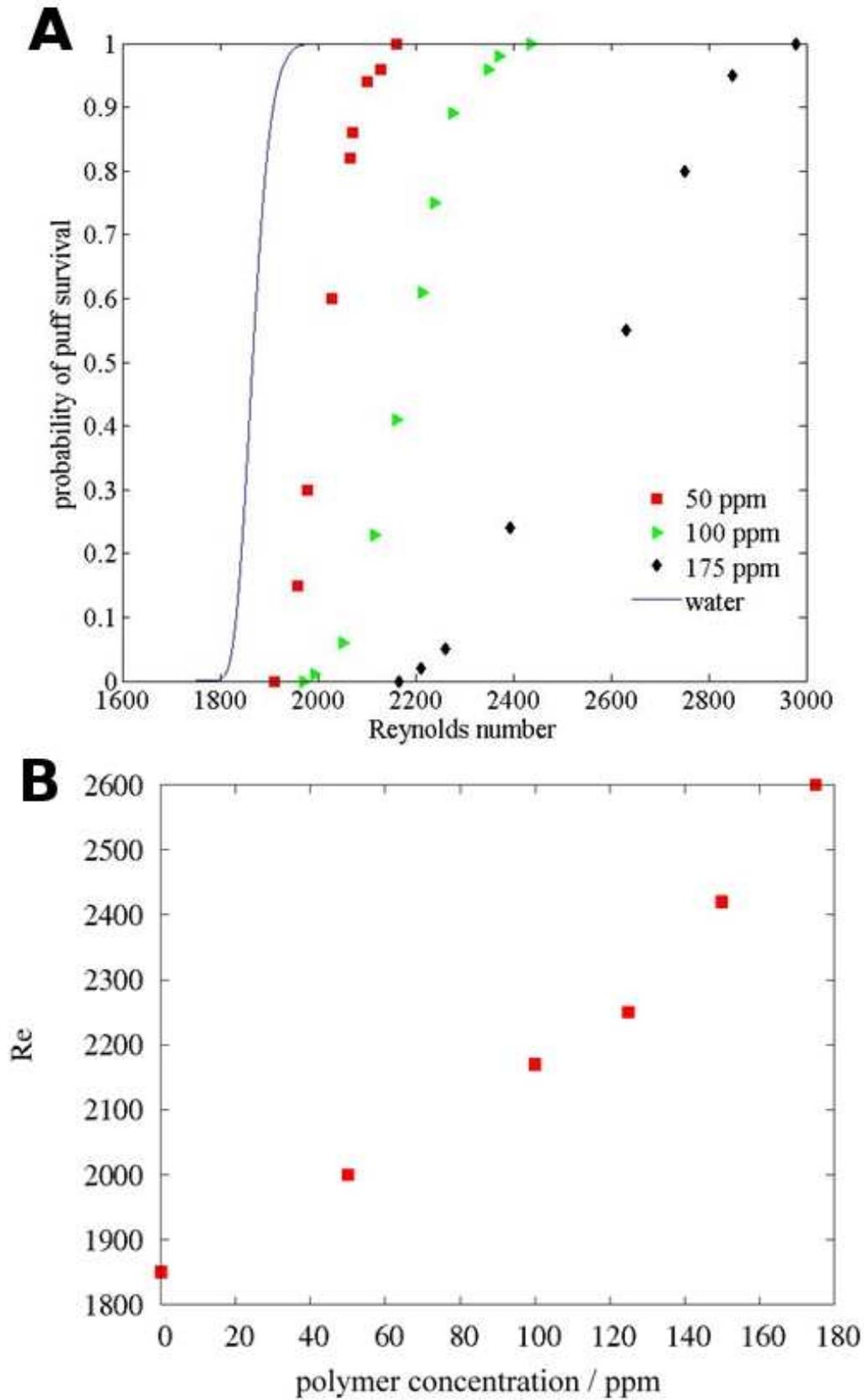

**Fig. 1**. Survival probability of turbulence. At the lowest Reynolds where turbulence can be observed in pipes it has the form of localized structures which decay following a memoryless process (16,19). The influence of polymers on this process is investigated for various concentrations. In order to create turbulence the laminar flow was perturbed by an impulsive injection (of the same fluid) for 20ms with sufficient amplitude to create a turbulent puff. The flow was then monitored at the exit of the pipe 760D downstream by means of visual

inspection using the same method as described in (16). While for laminar flow the fluid exits the pipe smoothly and follows a parabolic path, turbulent fluid (resulting from the different velocity profile) will exit the pipe at a different angle causing a downward deflection and temporary distortion of the out flowing jet. By continuously monitoring the outflow the survival probability of turbulent puffs was determined as a function of Re. (A) For increasing polymer concentrations survival probabilities decrease and the survival probability found in the Newtonian case is only recovered at larger Re. With increasing polymer concentration the typical S-shape of the probability distributions becomes more pronounced. (B) the Reynolds numbers with a 50% survival probability at the 760D observation point as a function of polymer concentration.

Surprisingly for polymer concentrations >=200ppm turbulent puffs could not be detected and instead a different type of disordered motion already sets in at lower Re: While in the Newtonian case turbulent fluctuations can first be sustained for Re~2000 (Fig. 2A, open squares) in a 500ppm solution disordered motion was observed for Re as low as 800 (Fig. 2B). Also in Newtonian fluids flows just above onset are intermittent (i.e. turbulent regions are interspersed by laminar ones (18), Fig. S2) in the polymer solutions fluctuations set in globally throughout the pipe (see Fig. S2). The instability observed in polymer solutions hence leads to a qualitatively different type of disordered motion, elasto-inertial turbulence. A further distinction between the two types of turbulence is that in the Newtonian case the onset is strongly hysteretic: unperturbed flows remain laminar up to large Re (in our setup to Re=6500, black squares in Fig. 2A,C) whereas perturbed flows display turbulence from around 2000. In contrast in a 500ppm solution perturbed and unperturbed flows become turbulent at the same Re (Fig. 2 B). Equally friction factors follow the same scaling and directly approach the maximum drag reduction asymptote (Fig. 2 D) without any excursions towards the Newtonian turbulence (so called Blasius) friction scaling. This observation suggests that the maximum drag reduction asymptote marks the characteristic drag of EIT, rather than being the consequence of an asymptotic adjustment of ordinary turbulence.

Further inspection shows that the elasto-inertial instability also appears for lower polymer concentrations (<200ppm). Here the instability sets in at larger Re and hence in the regime where in the presence of finite amplitude perturbations flows already exhibit Newtonian like (i.e. hysteretic and intermittent) turbulence. Starting from laminar flow without additional perturbations we find that with increasing Re these more dilute solutions will unavoidably turn turbulent at Reynolds numbers distinctly below the natural transition point (Re=6500) of this pipe as shown for a 100ppm solution in Fig. 2A and C (solid triangles).

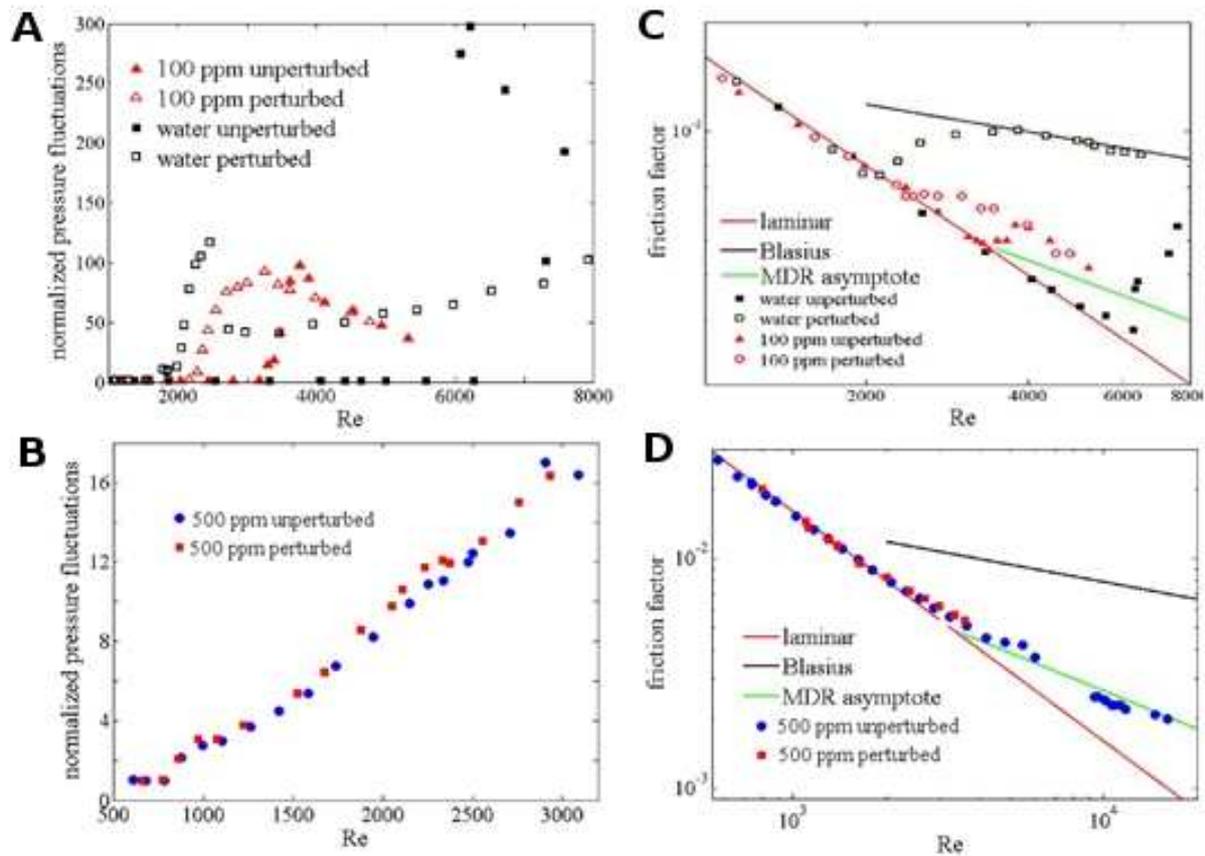

**Fig. 2.** Stability and friction scaling. For the Newtonian case (black symbols in A) turbulence can first be triggered by perturbations for Re of about 2000 where fluctuations increase rapidly(A). The quantity plotted is the fluctuation in pressure which was measured differentially between two pressure taps (1mm holes in the pipe wall) separated by 3D in the streamwise direction and located approximately 250D from the pipe exit. For a 100ppm (red data points in B) polymer solution turbulence cannot be triggered below Re 2200 (transition delay). At Re=3200 however an instability occurs (even in the absence of perturbations) this instability is solely caused by the presence of the polymers. B At higher polymer concentrations (here 500ppm) this instability occurs at much lower Re and the hysteresis typical for Newtonian turbulence has disappeared (A and C). The flow already becomes unstable at Re=800 (regardless of the presence of additional perturbations). From here (with increasing Re) the flow directly approaches the MDR friction scaling (D).

In contrast to the higher concentrations here the flow is intermittent consisting of localized turbulent regions (i.e. puffs) interspersed by non laminar, weakly fluctuating regions. As Re is further increased the spatial intermittency disappears and gives way to a uniformly fluctuating state and the friction values approach the MDR asymptote. The onset of instability is plotted in Fig. 3A as function of polymer concentration. Above the red curve the flow has become unstable and the friction factor begins to approach the MDR asymptote. The green data points mark the appearance of turbulent puffs shown in Fig. 1A (i.e. the threshold where the puff half-lifetime exceeds t=760). For parameter settings between the red and green data sets ordinary turbulence can be triggered by finite amplitude perturbations and the flow is hence hysteretic. Upon further increase of Re, once the red curve is crossed the flow will become unstable regardless of initial conditions.

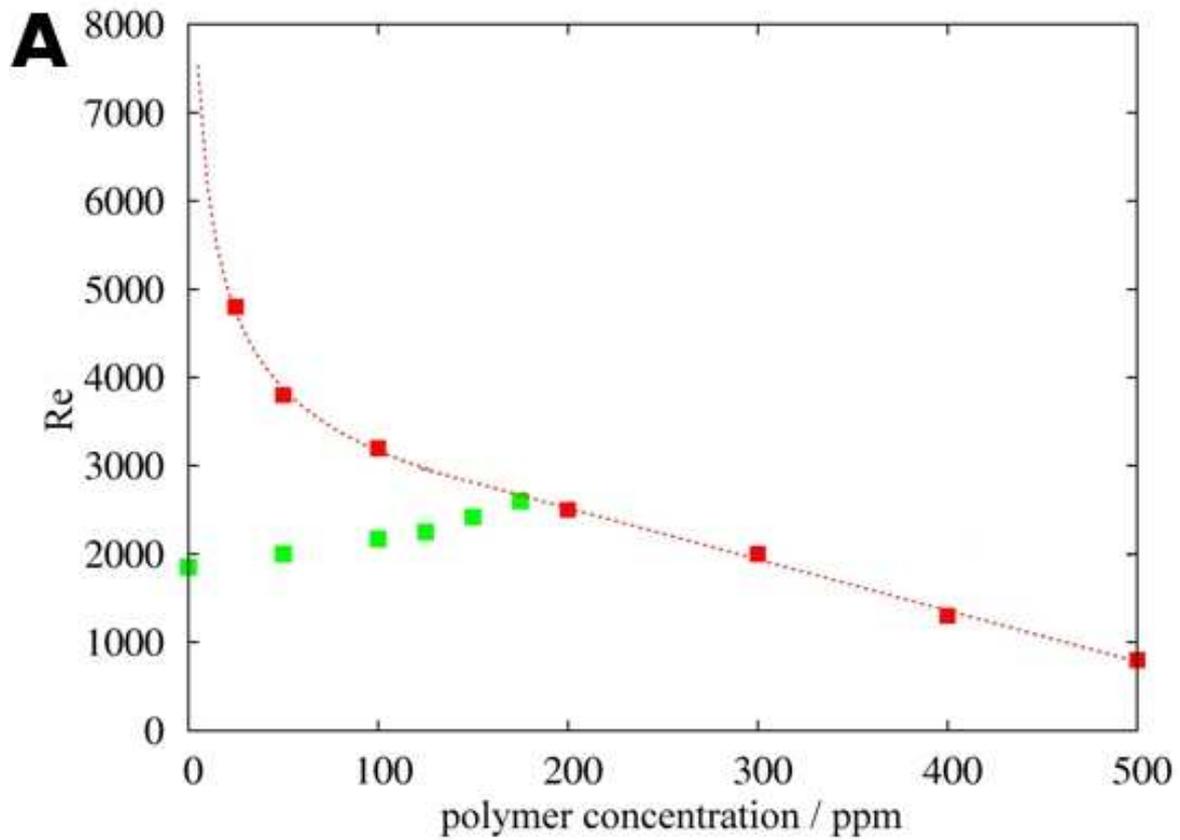

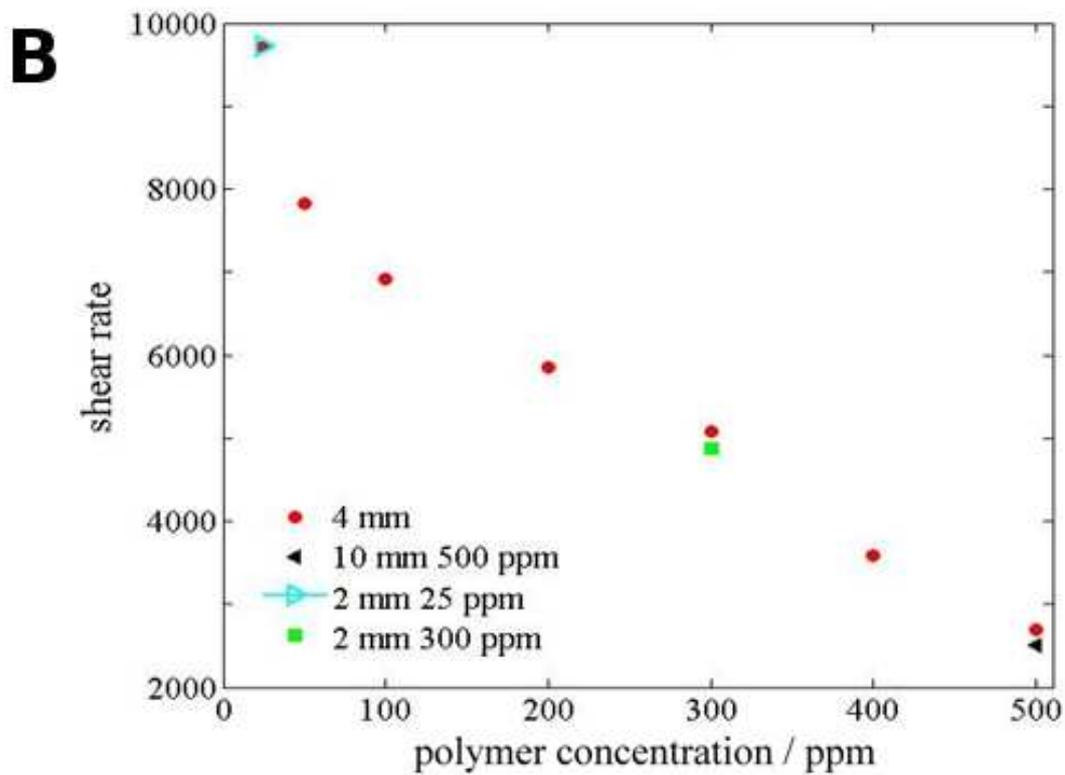

**Fig. 3.** The transition threshold to elasto-inertial turbulence is plotted for different concentrations (red squares). The red line is a guide to the eye. The green points mark the transition delay to ordinary turbulence (Fig. 1b). Consequently for concentrations below 200ppm the elasto-inertial instability sets in at Reynolds numbers where ordinary turbulence can already occur whereas in the D=4mm pipe above 200ppm only the elasto-inertial instability is found. (B) The red symbols marc the critical shear rate for the onset of EIT in the 4mm pipe (same

as red symbols in A). In addition critical shear rates were determined in a D=10mm and a D=2mm pipe. In these cases transition occurs at the same critical shear rate. Hence unlike ordinary turbulence the onset of EIT is not governed by Re but instead by the shear rate.

Finally experiments were carried out in pipes of diameters D=2mm and D=10mm (blue and red data points in Fig. 3B). When plotting the stability thresholds observed in the three pipes in terms of shear rate versus polymer concentration all data sets collapse. The latter observation shows that the elasto-inertial instability scales with the shear rate and not with Re (12). Hence in larger diameter tubes the instability will occur at large Re and typically be obscured by Newtonian turbulence. Inversely on micro scales this instability will occur at very low Re opening new avenues for mixing in micro-fluidic devices.

To gain further insights into the nature of the EIT, we conducted direct numerical simulation (DNS) of channel flow for a non-Newtonian fluid employing a constitutive model extensively used in the simulation of polymer drag reduction. The numerical methods and viscoelastic parameters are identical to those used in simulations of maximum drag reduction (20,21) (see SI for details). Great care was taken to resolve all flow scales relevant to the dynamics of such complex fluids requiring spatial and temporal resolutions significantly larger than for Newtonian turbulence. Each simulation is initially perturbed in such way that transition in Newtonian flow occurs at Re=6000, based on the bulk velocity and channel height. In qualitative agreement with the experiments we find that an instability develops at much lower *Re* in polymeric flows which again directly leads to the MDR asymptote, as shown in Fig. 4a for *Re*=1000. Whereas the corresponding Newtonian flow is perfectly laminar, Fig 4b shows fluctuations of wall pressure and significant chaos in the dynamics of polymer extension in the near-wall region.

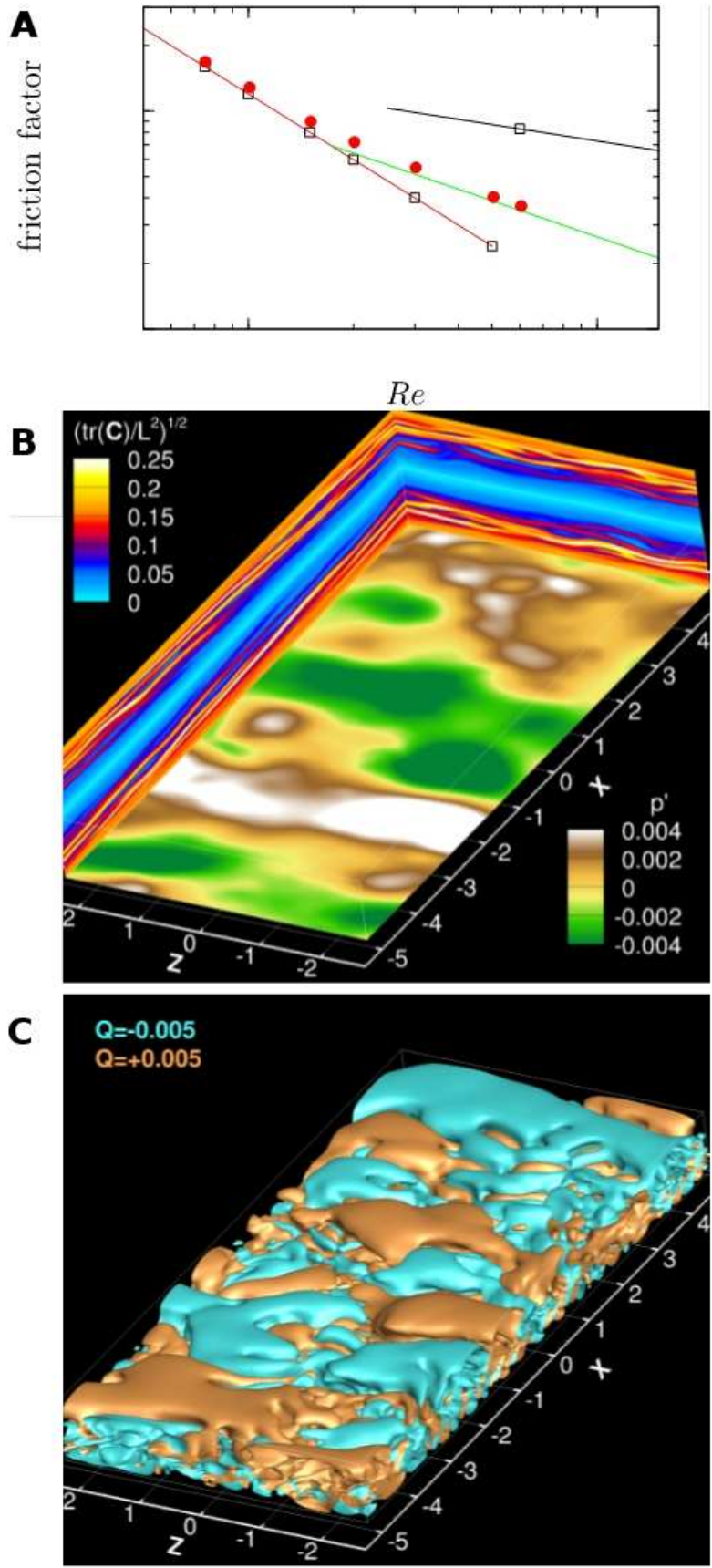

**Fig. 4.** Numerical simulation of elasto-inertial turbulence in a channel flow. A The red, green and black lines respectively highlight the laminar, turbulent and MDR distributions of the friction factor as a function of the Reynolds number based on the bulk velocity and the height of the channel. The simulations are performed in a channel flow of large transversal dimensions with periodic boundary conditions in horizontal dimensions. At time t=0, a perturbation is introduced in the form of space and time oscillations of blowing and suction at the walls for a fixed, short duration. The intensity of the perturbations is tailored so that transition is triggered at Re=6000 for the simulated water flow. Using the same perturbation, the simulated polymeric channel flow already shows departure from purely laminar flow around Re=750 (closed triangles). (B) Contours of pressure fluctuations on the bottom wall and polymer stretch in vertical planes (Re=750). Figure (C) shows isosurfaces of regions of slightly rotational (orange) or extensional (cyan) nature (Re=750), as identified by the second invariant of the velocity tensor $Q$ (see Supplementary Information for details).

**Instability Mechanism.** Closer inspection of the numerical data at the lowest simulated Reynolds number, Re=1000, reveals (see figure 4B) an interesting topological structure of EIT. Even though the flow is dominated by the mean shear, polymers are extended (large values of tr(C) /$L^2$) in sheet-like regions of large streamwise (x) and spanwise (z) dimensions. The sheets are stretched at an upward angle from the streamwise direction, indicative of extensional flow topology. These sheets also produce larger polymer extension than the surrounding mean shear does; an increase of the effective flow viscosity, through extensional viscosity, is therefore confined to these very sheets. The formation of sheets is driven by the convective transport of the polymer conformation tensor (see Eq. 3) . The response of the flow is observed in pressure fluctuations, shown in contours of wall pressure on the bottom wall of Fig. 4B.

The interactions between the sheet-like structure of polymer dynamics and pressure is best described by taking the divergence of the momentum transport of viscoelastic flow, yielding an elliptical equation for the pressure

$$\nabla^2 p = 2Q - \frac{1-\beta}{Re} \nabla \cdot (\nabla \cdot \mathbf{T}) \qquad [1]$$

where $Q = -\nabla \cdot ((\mathbf{u} \cdot \nabla) \cdot \mathbf{u})$ is the second variant of the velocity gradient tensor and also a measure of the local flow topology (25). Note that $Q$ is also the difference between the square norm of the rotation rate and deformation rate tensors. Figure 4C shows isosurfaces of positive and negative $Q$. As shown the flow is structured in alternating regions of rotation ($Q>0$) and deformation ($Q<0$) which are aligned in the spanwise direction. The cylindrical structure of these regions is attributed to the elliptical character of Eq. (1).

Underlying EIT is hence a self-sustaining cycle where small velocity perturbations cause the formation of sheets of extended polymers through convective transport. The flow response, through pressure, sustains velocity fluctuations, therefore closing the cycle.

It is noteworthy that the key elements of the mechanism of EIT (nonlinear advection of stress, stretching by flow and flow response via pressure) are common features to many viscoelastic fluids. The only viscoelastic requirement is that extensional viscosity increases in elongational flows, which for example is also a property of surfactant additives (26). Additional experiments confirmed that also in surfactant solutions, as in the polymer case fluctuation levels increase at Re lower than for a Newtonian fluid (see Fig. S8), again indicating the onset of an instability.

In summary we have identified an instability for dilute polymer solutions that gives rise to the well known empirical friction law, the so called maximum drag reduction asymptote. The MDR asymptote is therefore not the asymptotic limit of ordinary turbulence weakened by polymeric action, but instead it is the characteristic friction behaviour of a different state of motion: elasto-inertial turbulence. Our observations infer that this type of fluid motion replaces ordinary turbulence and dominates the dynamics in elastic fluids at sufficiently large shear rates.

## Methods.

**Experimental Methods.** Experiments were carried out in a pipe made of ~1.2m long precision bore segments with an inner diameter of D=4mm+/-0.01and a total length of about L/D=900. The flow was gravity driven and the fluid temperature controlled so that the flow rate could be held constant to typically within +/-0.2% (details of a similar setup can be found in (15)). The sample solutions were either pure water or different amounts of polyacrylamide with a molecular weight of $5 \times 10^6$ amu (PAAm, Sigma-Aldrich, Munich, Germany) in water. The shear viscosity increased linearly with the polymer concentration and no shear thinning was observed but a pronounced elastic behavior was found in the elongational flow of a Capillary Break-up Elongational Rheometer (CaBER). The rheological characterization is given in the SI. A carefully designed inlet of the pipe allowed to keep flows of pure water laminar up to Re~6500 (natural transition point for this pipe). Here the Reynolds number is defined as Re=UD/ν, where U is the mean flow speed and ν the kinematic viscosity. While laminar (Newtonian) pipe flow is stable for all Re, turbulence of appreciable lifetime can be triggered by perturbations of finite amplitude once Re approaches 2000 (16,17). In the present set up turbulence was triggered by injecting fluid through a small hole in the pipe wall situated 140D from the inlet, or alternatively for continuous triggering of turbulence an obstacle (a ~2 cm long, ~1mm thick twisted wire) could be placed downstream of the inlet.

**Numerical Methods.** The flow is governed by the incompressible Navier Stokes equations with the addition of a viscoelastic stress

$$\partial_t \mathbf{u} + (\mathbf{u} \cdot \nabla)\mathbf{u} = -\nabla p + \frac{\beta}{\mathrm{Re}} \nabla^2 \mathbf{u} + \frac{1-\beta}{\mathrm{Re}} \nabla \cdot \mathbf{T} \text{ and } \nabla \cdot \mathbf{u} = \mathbf{0} \qquad 1]$$

in a rectangular domain with periodic boundary conditions in the streamwise and spanwise directions and no-slip at the walls. where u is the velocity vector, p the pressure and Re the Reynolds number. The flow is driven by a bulk force to maintain a constant mass flow rate. The velocity and length scales used to form the Reynolds number and to normalize the flow variables are the bulk velocity and the height of the channel, respectively. The polymer stress tensor T in Eq. (1) is derived from the following transport equation

$$\partial_t \mathbf{C} + (\mathbf{u} \cdot \nabla)\mathbf{C} = \mathbf{C} \cdot \nabla \mathbf{u} + \nabla \mathbf{u}^T \cdot \mathbf{C} - \mathbf{T} \text{ with } \mathbf{T} = \frac{1}{\mathrm{Wi}} \left( f(\mathbf{C})\mathbf{C} - \mathbf{I} \right) \text{ and } f(\mathbf{C}) = \frac{1}{1 - \mathrm{trace}(\mathbf{C})/L^2} \quad 2]$$

where C is the conformation tensor and *f* is the Peterlin function based on *L* the upper limit of polymer extension. The polymer solution is characterized by the Weissenberg number Wi, which is the ratio of polymer relaxation time to flow scale, here the inverse of the wall shear. In Eq. (1), the coefficient is the ratio of the zero-shear viscosity of the polymer solution to the solvent viscosity. The numerical method used to solve Eqs. (1) and (2) is described in (20). The computational domain dimensions and resolution are 10HxHx5H and 256x161x256, respectively. Doubling the dimensions and resolution in transversal directions did not yield appreciable change in statistics nor increasing the resolution in the wall-normal direction.